\documentclass[aps,prb,twocolumn,superscriptaddress]{revtex4-2}
\usepackage{graphicx}
\usepackage{caption}
\usepackage{latexsym}
\usepackage{amssymb}
\usepackage{amsfonts}
\usepackage[vcentermath]{youngtab}
\usepackage{upgreek}
\usepackage{bm,dsfont}
\usepackage{multirow}
\usepackage{enumitem}
\usepackage{color}
\usepackage{hyperref}
\usepackage{float}
\hypersetup{
colorlinks = true,
linkcolor = [rgb]{0.70,0.13,0.13},
citecolor = [rgb]{0.13,0.55,0.13},
urlcolor = [rgb]{0.25, 0.41, 0.88}}
\usepackage{makecell}

\newcommand{\dd}{\mathrm{d}}

\newcommand{\ii}{\mathrm{i}}

\newcommand{\E}{\mathop{\mathbb{E}}}

\newcommand{\SU}{\mathrm{SU}}

\newcommand{\SO}{\mathrm{SO}}

\newcommand{\su}{\mathfrak{su}}

\newcommand{\scL}{\mathcal{L}}

\newcommand{\scR}{\mathcal{R}}

\newcommand{\vect}[1]{{\bm{#1}}}

\newcommand{\eq}[1]{\begin{equation}#1\end{equation}}
\newcommand{\eqs}[1]{\begin{equation}\begin{split}#1\end{split}\end{equation}}
\newcommand{\eqnref}[1]{Eq.\,\eqref{#1}}
\newcommand{\figref}[1]{Fig.\,\ref{#1}}
\newcommand{\tabref}[1]{Tab.\,\ref{#1}}
\newcommand{\secref}[1]{Sec.\,\ref{#1}}
\newcommand{\appref}[1]{Appendix\,\ref{#1}}

\usepackage{amsmath}
\usepackage{subfig}

\begin{document}

\title{Machine Learning Symmetry Discovery for Integrable Hamiltonian Dynamics}
\author{Wanda Hou}
\affiliation{Department of Physics, University of California at San Diego, La Jolla, CA 92093, USA}
\author{Molan Li}
\affiliation{Department of Physics, University of California at San Diego, La Jolla, CA 92093, USA}
\author{Yi-Zhuang You}
\email{yzyou@physics.ucsd.edu}
\affiliation{Department of Physics, University of California at San Diego, La Jolla, CA 92093, USA}
\date{\today}

\begin{abstract}
We propose a data-driven Machine-Learning Symmetry Discovery (MLSD) framework for identifying continuous symmetry generators and their Lie-algebraic structure directly from phase-space trajectory data expressed in canonical coordinates. MLSD parameterizes candidate conserved quantities with neural networks and learns antisymmetric structure coefficients by enforcing Poisson-bracket closure, supplemented by a weak independence regularizer. We validate MLSD on two integrable benchmark systems—the three-dimensional Kepler problem and the three-dimensional isotropic harmonic oscillator—recovering the expected non-Abelian algebras (respectively $\mathfrak{so}(4)$ and $\mathfrak{su}(3)$) up to basis transformations. This work focuses on integrable benchmark dynamics, where global conserved quantities are well-defined and admit compact representations learnable from canonical-coordinate trajectories. Extending symmetry discovery to mixed or chaotic phase-space regimes is an important direction for future work.
\end{abstract}

\maketitle

\section{Introduction}

The term \textit{symmetry} originates from the Greek words \textit{syn} (meaning ``same'') and \textit{metron} (meaning ``measure''). At its core, symmetry reflects a fundamental concept in our understanding of the natural world: the ability to preserve identical measurable properties under transformations or dynamics. Much of the progress in physics as a science has come from uncovering symmetries within physical systems, deepening our insight into the laws governing the universe\cite{2022Symm...14.1475W,PhysRevLett.122.175701,2017NatCo...8...50P,PhysRevB.87.155114}.

While the discovery of symmetries is invaluable, it is often a challenging task that typically requires the sophisticated expertise of physicists. To address this challenge, advancing field of machine learning\cite{lecun2015deep,2019RvMP...91d5002C,2016arXiv160207576C} offers a promising set of tools for assisting symmetry discovery from data in controlled physical settings. One significant advantage of machine learning in this context is its ability to learn and compress from large volumes of data. This data-driven approach provides a systematic way to extract patterns and correlations from data that may not be evident from standard theoretical analysis. By harnessing the vast amounts of data generated in scientific research, machine learning holds the potential to accelerate discovery and provide new insights into the fundamental laws and symmetries governing physical systems\cite{2023Natur.620...47W,kashinath2021physics,2020NatPh..16.1050H}.

Previous work has demonstrated learning-based approaches to uncovering continuous symmetries in classical mechanical systems, primarily in settings with regular dynamics where conserved quantities admit stable functional representations\cite{2023arXiv231000105Y,PhysRevLett.128.180201,PhysRevLett.126.180604,pmlr-v202-yang23n,otto2023unified,forestano2023deep,2024arXiv241008087V,PhysRevE.106.045307}. In this paper we similarly focus on benchmark Hamiltonian systems with integrable dynamics, where the relevant conserved quantities are well-defined on the sampled phase-space region and admit compact representations. A natural next step within this line of work is to develop a systematic framework for discovering continuous Lie group symmetries, particularly non-Abelian ones, directly from trajectory data, without assuming prior analytic knowledge of the conserved quantities.

In this study, we propose a Machine-Learning Symmetry Discovery (MLSD) algorithm designed to identify continuous Lie group symmetries in classical mechanical systems with integrable Hamiltonian dynamics, using simulated time-evolution trajectory data. The MLSD algorithm identifies conservation laws that correspond to continuous symmetry transformations. After training, MLSD outputs a three-way tensor  $f$, representing the Lie algebra structure coefficients of the discovered set of symmetry transformations. The structure coefficients provide a basis-independent representation that allows MLSD to identify both Abelian and non-Abelian symmetry algebras within the considered setting.

Within this deliberately scoped benchmark setting, MLSD provides an end-to-end pipeline that (i) learns candidate conserved quantities from trajectories, (ii) recovers a basis-independent Lie-algebra signature via structure coefficients and Killing-form analysis, and (iii) reports stability and reproducibility diagnostics suitable for systematic comparison across symmetry dimensions.

The paper is organized as follows: we first review the definition of continuous symmetries in classical systems, then explain each loss term designed in MLSD to discover symmetries using neural networks, and finally examine our proposed algorithm on two tasks: discovering $\SO(4)$ symmetry in the three-dimensional Kepler problem; and discovering $\SU(3)$ symmetry in the three-dimensional harmonic oscillator.

\begin{figure*}
\begin{center}
    \includegraphics[width=380pt]{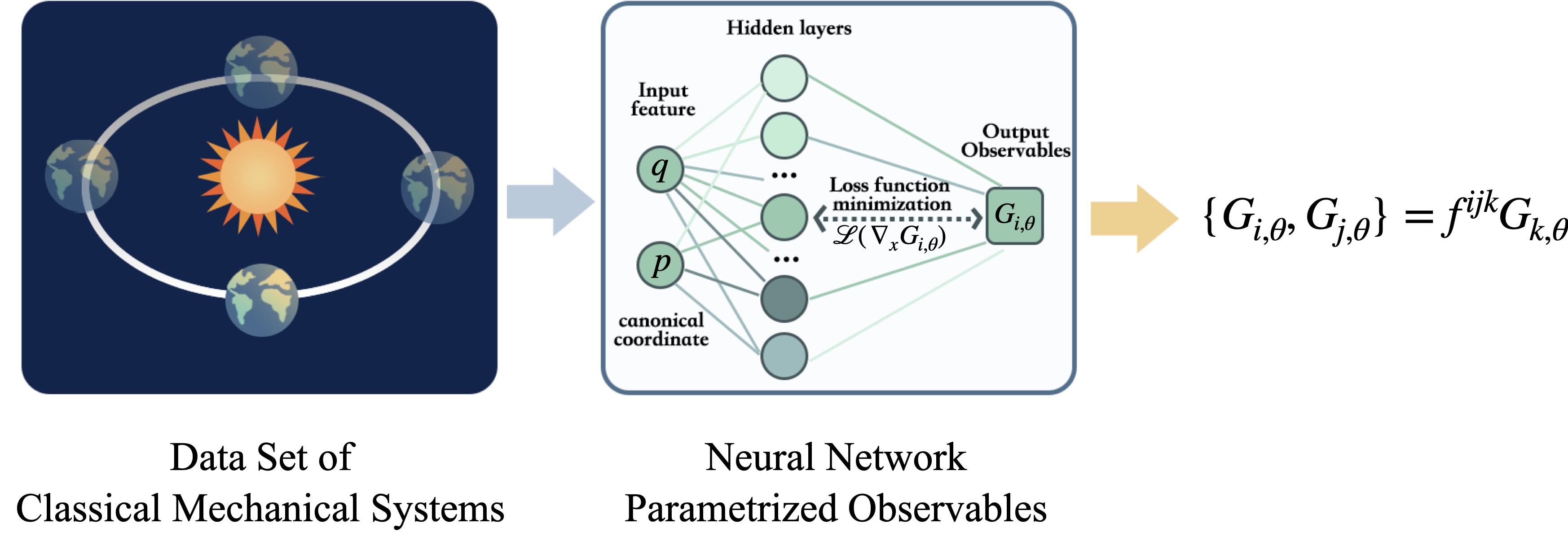}
\end{center}
\caption{The Machine Learning Symmetry Discovery (MLSD) framework takes time evolution data from a classical system and feeds the canonical coordinates into neural networks to predict a physical quantity. By optimizing the loss function, the predicted observables converge to conserved quantities, and the corresponding structure coefficients reveal the underlying symmetry group.}
\label{fig:pip}
\end{figure*}

\section{Methodology}\label{sec: method}

Our goal of symmetry discovery differs from those explored in previous studies. For example, \cite{2023arXiv231000105Y,pmlr-v202-yang23n} focuses on identifying transformations that preserve time evolution trajectories, while \cite{PhysRevLett.128.180201,PhysRevLett.126.180604} examines whether specific proposed transformations can be learned by neural networks. In contrast, our approach seeks to uncover symmetry transformations that extend beyond preserving the shape of trajectories and instead focus solely on preserving energy, as similarly discussed in \cite{otto2023unified,forestano2023deep,2024arXiv241008087V}, without assuming analytic forms for the conserved quantities. In this section, we begin by reviewing the concept of symmetry in classical systems as it pertains to our study. We then introduce the architecture of the Machine Learning Symmetry Discovery (MLSD) framework, which is designed to identify these symmetries.

\subsection{Continuous symmetry and symmetry group of a physical system}\label{sec: sym}

The time evolution dynamics of a physical system is encoded in the Hamiltonian $H(\vect{x})$, which depends on the canonical coordinates $\vect{x}:=(\vect{q}, \vect{p})$, with $\vect{q}$ and $\vect{p}$ being the conjugate position and momentum variables. A continuous symmetry of the system is defined by the invariance of the Hamiltonian under a continuous family of canonical transformations generated by a conserved quantity. This can be expressed through the vanishing Poisson bracket between $H(\vect{x})$ and the conserved quantity $G_i(\vect{x})$, where $i\in\{1,2,...,n\}$ and $n$ denote the dimension of the symmetry group. The identification of the continuous symmetry group (Lie group) follows directly from the Lie algebra structure coefficients $f^{ijk}$, which are obtained from the commutators between the quantities.
\eq{\label{eq: sym1}
\{H,G_i\}=0,\forall\ i\in\{1,2,...,n\}.
}
\eq{\label{eq: sym2}
\{G_i,G_j\}=\sum_{k}f^{ijk}G_k,\forall\ i,j\in\{1,2,...,n\}.
}
Here the Poisson bracket $\{A,B\}$ between two canonical functions $A(\vect{x})$ and $B(\vect{x})$ is defined by the following vector-matrix-vector multiplication
\eq{\{A,B\}:=(\nabla_\vect{x}A)^\intercal \mathbf{J} (\nabla_\vect{x}B),}
where $\nabla_\vect{x}$ is the gradient operator, and the matrix
\eq{\label{eq: def J}\mathbf{J}= \begin{pmatrix}  & \mathds{1}_{d\times d} \\ -\mathds{1}_{d\times d} &   \end{pmatrix}}
characterizes the symplectic metric of the canonical coordinate  $\vect{x}$ in the phase space.

Since the three-way tensor $f^{ijk}$ is anti-symmetric by definition, we can assign $G_0(\vect{x}):=H(\vect{x})$ together with $f^{0jk}=f^{i0k}=f^{ij0}=0$. Therefore \eqnref{eq: sym1} and \eqnref{eq: sym2} can be combined for compactness as
\eq{\label{eq: sym3}
\{G_i,G_j\}=\sum_{k}f^{ijk}G_k,\forall\ i,j\in\{0,1,2,...,n\}.
}
The goal in this work is to discover a set of independent conserved quantities $G_i(\vect{x})$ and associated structure coefficients $f^{ijk}$ that describe the symmetry algebra, within the expressivity of the chosen parametrization and the phase-space coverage of the dataset. The special $G_0$ component corresponds to the Hamiltonian $H$, relating to energy conservation. Its dependence $H(\vect{x})$ on $\vect{x}$ can be learned from a series of time evolution data $\vect{x}(t)$, assuming the underlying principle of Hamiltonian dynamics of classical mechanics.

\subsection{Learning continuous symmetry transformations}\label{sec: loss}

Considering a classical system in spatial dimension $d$, let
$\vect{X}=\{\vect{x}(t) \in \mathbb{R}^{2d} : \dot{\vect{x}}=\{\vect{x},H\}\}$
denote a dataset of time-series samples, where each sample is a trajectory of
the canonical coordinates $\vect{x}(t)$ evolving under Hamiltonian dynamics
generated by an unknown Hamiltonian $H$.
Throughout this work, we assume that the dataset $\vect{X}$ consists of
simulated trajectories expressed in canonical coordinates.
We prepare the dataset $\vect{X}$ to contain multiple trajectories initialized
from diverse regions of phase space, so that the training procedure probes the
global structure of the dynamics rather than a single dynamical regime.

The goal is to uncover the continuous Lie-group symmetries of such a classical
dynamical system as defined by \eqnref{eq: sym3}.
To this end, we propose the Machine-Learning Symmetry Discovery (MLSD)
algorithm, which uses neural networks (NNs) to parametrize each conserved
quantity in \eqnref{eq: sym3} as $G_{i,\theta}$, where $\theta$ denotes the
optimizable parameters of the neural networks.

The \textit{Noether’s theorem} states that these conserved quantities $G_{i,\theta}$ are also symmetry generators of the corresponding continuous symmetry. An infinitesimal symmetry transformation of $\vect{x}$ is given by the gradient of $G_{i,\theta}$ with respect to $\vect{x}$:
\eqs{\label{eq: tran}
\vect{x}'=& \vect{x}+\{\vect{x},G_{i,\theta}\}\epsilon+\mathcal{O}(\epsilon^2)
\\=& \vect{x}+\mathbf{J}(\nabla_{\vect{x}} G_{i,\theta})\epsilon+\mathcal{O}(\epsilon^2).
}
where $\epsilon$ denotes an infinitesimal variation and $\mathbf{J}$ is the symplectic metric defined in \eqnref{eq: def J}. Based on this principle, we can parametrize a conserved quantity  $G_{i,\theta}$  and use its gradient, $\nabla_{\vect{x}} G_{i,\theta}$, to generate a symmetry transformation $\delta \vect{x}=\mathbf{J}(\nabla_\vect{x}G_i)\epsilon$. In contrast to prior studies that directly model the symmetry transformation $\delta \vect{x}$ by neural networks\cite{2023arXiv231000105Y,PhysRevLett.128.180201,PhysRevLett.126.180604,pmlr-v202-yang23n,forestano2023deep}, our approach of modeling $\delta\vect{x}$ indirectly through gradients of scalar functions $G_{i,\theta}$ is simpler, while inherently respecting the curl-free constraint $\nabla_{\vect{x}} \times (\nabla_{\vect{x}} G_{i,\theta}) = 0$ that should otherwise be imposed on $\delta \vect{x}$ as well. We remark that MLSD assumes access to trajectories expressed in canonical coordinates $(\mathbf{q},\mathbf{p})$, for which the symplectic form has the standard Darboux representation. Handling data of arbitrary coordinates would require learning the symplectic structure or a canonical transformation, which is beyond the scope of this work.


\begin{itemize}
\item Hamiltonian learning

Let us start from learning the Hamiltonian of the classical system. Using \eqref{eq: tran}, we expect $G_{0,\theta}$ can generate one step time evolution by $\{\vect{x},G_{0,\theta}\}=\dot{\vect{x}}$. Hence, the Hamiltonian of the sytem can be learned by minimizing the mean-square loss function:
\eq{\label{eq: lossH}
\mathcal{L}_{\text{H}}=\E_{\vect{x}(t)\in\vect{X}}\int\big\Vert\mathbf{J}\nabla_{\vect{x}} G_{0,\theta}-\dot{\vect{x}}\big\Vert^2\dd t.
}
where the trajectory $\vect{x}(t)$ is sampled over the time series dataset $\vect{X}$. In practice, the time derivative is approximated by $\dot{\vect{x}}(t)\simeq \frac{\vect{x}(t+\epsilon)-\vect{x}(t-\epsilon)}{2\epsilon}$ along the trajectory for some small $\epsilon$. It should be understood that the integrand is time dependent along the trajectory $\vect{x}(t)$, and the time $t$ integrates over the trajectory sampled from the dataset $\vect{X}$.

\end{itemize}

\begin{itemize}
\item Symmetry discovery

The symmetry discovery process begins by inputting a conjectured dimension $n$ of the Lie group. Therefore, $n+1$ individual neural networks are used to predict each conserved quantity. The structure coefficient $f^{ijk}$ is parametrized to be anti-symmetric, defined as $f^{ijk}=\eta^{ijk}-\eta^{ikj}+\eta^{jki}-\eta^{jik}+\eta^{kij}-\eta^{kji}$, for all $i,j,k\neq 0,
$ with the parameter $\eta$ being optimizable. In order to recognize the symmetry group formed by all $G_{i,\theta}$, we train on the dataset $\vect{X}$ with the following mean square loss:
\eqs{\label{eq: lossG}
\mathcal{L}_{\text{G}}=\E_{\vect{x}(t)\in\vect{X}} \frac{1}{(n+1)^2}\sum_{i,j=0}^{n}\int\Big\Vert&(\nabla_{\vect{x}} G_{i,\theta})^{\intercal}\mathbf{J}(\nabla_{\vect{x}} G_{j,\theta})\\
&-\sum_{k=0}^{n}f^{ijk}G_{k,\theta}\Big\Vert^2\dd t.
}
It should be understood that the integrand is evaluated instantaneously at each time $t$ and then integrated over  the trajectory $\vect{x}(t)$ sampled from the dataset $\vect{X}$.

\end{itemize}

In addition to $\mathcal{L}_{\text{H}}$ and $\mathcal{L}_{\text{G}}$, we introduce a third term in the loss function to encourage linear independence among the learned symmetry transformations. Without such a term, the optimization may converge to trivial solutions in which several $G_{i,\theta}$ correspond to equivalent conserved quantities. To quantify independence, we construct the $(n+1)\times 2d$ matrix
\begin{equation}
\mathbf{M}(\vect{x})
=
\big(\mathbf{J}\nabla_{\vect{x}} G_{0,\theta},
      \nabla_{\vect{x}} G_{1,\theta},
      \dots,
      \nabla_{\vect{x}} G_{n,\theta}\big)^\intercal,
\end{equation}
and consider the eigenvalues $\mu^{i}(\vect{x}) \ge 0$ ($i=1,\dots,n+1$) of the Gram matrix $\mathbf{M}(\vect{x})\mathbf{M}^{\intercal}(\vect{x})$. We then define normalized eigenvalues
\begin{equation}
\tilde{\lambda}^{i}(\vect{x})
=
\frac{\mu^{i}(\vect{x})}{\sum_{j=1}^{n+1} \mu^{j}(\vect{x})},
\end{equation}
which satisfy $\tilde{\lambda}^{i}(\vect{x}) \in [0,1]$ and $\sum_{i=1}^{n+1} \tilde{\lambda}^{i}(\vect{x}) = 1$. These normalized eigenvalues describe how the learned generators distribute their “weight’’ across independent directions in the space spanned by $\{\nabla_{\vect{x}} G_{i,\theta}\}$.

As an independence-promoting objective, we use the Shannon entropy of the normalized eigenvalues,
\begin{equation}
\mathcal{S}_{\mathrm{ind}}(\vect{x})
=
- \sum_{i=1}^{n+1}
\tilde{\lambda}^{i}(\vect{x})
\log\!\big(\tilde{\lambda}^{i}(\vect{x})+\varepsilon_\lambda\big),
\end{equation}
where a small constant $\varepsilon_\lambda$ is added inside the logarithm for numerical stability.
In all experiments we set $\varepsilon_\lambda = 10^{-5}$.
Since the eigenvalues are normalized, this entropy is scale-invariant and remains well-behaved
when some $\tilde{\lambda}^i$ approach zero.
A higher value of $\mathcal{S}_{\mathrm{ind}}(\vect{x})$ corresponds to a more isotropic eigenvalue
spectrum, indicating that the learned generators span multiple independent directions, whereas a
lower entropy signals collapse onto a lower-dimensional subspace.
The corresponding independence term in the loss function is
\begin{equation}
\label{eq: lossI}
\mathcal{L}_{\text{I}}
=
- \E_{\vect{x}(t)\in\vect{X}}
\int \mathcal{S}_{\mathrm{ind}}(\vect{x}(t)) \,\dd t.
\end{equation}
In practice, $\mathcal{L}_{\text{I}}$ is weighted by a small factor $\beta \ll 1$ in the total loss,
so that it acts as a weak regularizer that discourages degeneracy among generators without
dominating the Hamiltonian-learning and algebra-closure objectives. Similar entropy-based independence regularization has also been explored in \cite{PhysRevE.106.045307}.

In summary, the overall loss function is
\eq{\label{eq: loss}
\mathcal{L}
=
\mathcal{L}_{\text{H}}
+
\alpha\,\mathcal{L}_{\text{G}}
+
\beta\,\mathcal{L}_{\text{I}} .
}
Throughout this work we set $\alpha = 1$, so that
$\mathcal{L}_{\text{H}}$ and $\mathcal{L}_{\text{G}}$ contribute at comparable magnitude,
while the independence term $\mathcal{L}_{\text{I}}$ is weakly weighted with
$\beta = 10^{-4}$.

After training, if the dataset \( \vect{X} \) corresponds to a system with a continuous
$n$-dimensional symmetry, the first two loss terms,
\eqnref{eq: lossH} and \eqnref{eq: lossG}, are expected to approach zero,
while the independence term \eqnref{eq: lossI} remains finite.
Conversely, if the system does not possess an $n$-dimensional symmetry,
the first two loss terms remain bounded away from zero.

In practice, even when the correct symmetry dimension is supplied,
$\mathcal{L}_{\text{H}}$ and $\mathcal{L}_{\text{G}}$ may not converge exactly to zero
due to the finite expressive capacity of the neural networks and optimization limitations.
Nevertheless, a clear separation in the converged loss values is observed between the
true symmetry dimension and mismatched (over- or under-parameterized) cases.
The specific symmetry group is then identified by extracting the structure coefficients
$f^{ijk}$, as detailed in \appref{app: killing}.

\begin{table*}[t]
  \centering
  \begin{tabular}{lcccc} 
     & Hamiltonian & Conserved quantities & Lie Algebra & Lie Group\\  \hline
    Harmonic oscillator & $\displaystyle H=\frac{\vect{p}^2}{2}+\frac{\vect{q}^2}{2}$ & \makecell{$\displaystyle G_i=\frac{1}{2}(\vect{q}+\ii\vect{p})^{T}M_{i}(\vect{q}-\ii\vect{p}),$ \\ $\displaystyle M_{i}\in \text{Gell-Mann matrices}.$}& \makecell{$\displaystyle \{G_i,G_j\}=\frac{1}{2}(\vect{q}+\ii\vect{p})^{T}[M_i,M_j](\vect{q}-\ii\vect{p}).$} & SU(3)\\
     \hline 
    Kepler problem  & $\displaystyle H=\frac{\vect{p}^2}{2}-\frac{1}{|\vect{q}|}$ & \makecell{$\displaystyle \vect{L}=\vect{q}\times\vect{p},$ \\ $\displaystyle \vect{A}=\vect{p}\times\vect{L}-\hat{q};$\\
    $\vect{\scL},\vect{\scR}=\tfrac{1}{2}(\vect{L}\pm\vect{A}/\sqrt{-2H}).$}  & \makecell{$\displaystyle \{\mathcal{L}_i,\mathcal{L}_j\}=\epsilon^{ijk}\mathcal{L}_k,$ \\ $\displaystyle \{\mathcal{R}_i,\mathcal{R}_j\}=\epsilon^{ijk}\mathcal{R}_k, $\\ $\displaystyle \{\mathcal{L}_i,\mathcal{R}_j\}=0.$} & SO(4)\\
     
  \end{tabular}
  \caption{Summary of hidden symmetries in harmonic oscillator and Kepler problem.}
  \label{tab:1}
\end{table*}

\section{Empirical results\label{sec: er}}

In this section, we apply MLSD to discover symmetries in the
\textit{harmonic oscillator} and the \textit{Kepler problem}, summarized in
\tabref{tab:1}.
A detailed analysis of the $\SU(3)$ symmetry of the harmonic oscillator and the
$\SO(4)$ symmetry of the Kepler problem is provided in \appref{app: sym}.
We demonstrate that MLSD can successfully identify the hidden symmetries in both
systems\cite{github}.
Additionally, in the harmonic oscillator task, we use a quadratic
parametrization to show that MLSD can explicitly reconstruct the eight
Gell--Mann matrices that form the $\su(3)$ Lie algebra.

Throughout this section, all trajectory rollouts are generated using a
symplectic integration scheme.
In our implementation, both the ground-truth Hamiltonians and the learned
Hamiltonian $G_{0,\theta}$ are parameterized in a separable form,
\begin{equation}
H(\mathbf{q},\mathbf{p}) = T(\mathbf{p}) + V(\mathbf{q}),
\end{equation}
which enables the use of the velocity-Verlet (leapfrog) symplectic integrator.
The integration scheme advances the canonical variables as
\begin{equation}
\begin{aligned}
\mathbf{p}_{n+\frac{1}{2}} &=
\mathbf{p}_n - \frac{h}{2}\nabla_{\mathbf{q}} V(\mathbf{q}_n), \\
\mathbf{q}_{n+1} &=
\mathbf{q}_n + h\,\nabla_{\mathbf{p}} T(\mathbf{p}_{n+\frac{1}{2}}), \\
\mathbf{p}_{n+1} &=
\mathbf{p}_{n+\frac{1}{2}} - \frac{h}{2}\nabla_{\mathbf{q}} V(\mathbf{q}_{n+1}).
\end{aligned}
\label{eq:velocity-verlet}
\end{equation}
where $h$ denotes the integration time step.
This integrator is symplectic by construction, as it corresponds to a
composition of exact Hamiltonian flows generated separately by
$T(\mathbf{p})$ and $V(\mathbf{q})$, and ensures reliable long-time energy
conservation for both the learned and ground-truth dynamics. Energy conservation is assessed by tracking the deviation
\begin{equation}
\Delta H(t_n) \equiv \big| H(\mathbf{q}_n,\mathbf{p}_n) - H(\mathbf{q}_0,\mathbf{p}_0) \big|,
\end{equation}
along each trajectory, which provides a direct measure of energy drift in the learned dynamics.

To quantify energy conservation beyond time-resolved drift curves, we also compute the
root-mean-square (RMS) energy deviation along each trajectory,
\begin{equation}
\Delta H_{\mathrm{RMS}}
=
\sqrt{\frac{1}{N}\sum_{n=1}^{N}
\big[ H(\mathbf{q}_n,\mathbf{p}_n) - H(\mathbf{q}_0,\mathbf{p}_0) \big]^2 },
\end{equation}
where $N$ denotes the number of integration steps.
The RMS deviation provides a single scalar measure that summarizes the typical magnitude of
energy drift accumulated over the trajectory.
For clarity, the RMS energy deviation for each trajectory is reported as an inset in the
corresponding energy-drift panels.

\subsection{Kepler problem}

\begin{figure}[htbp]
\begin{center}
\includegraphics[width=240pt]{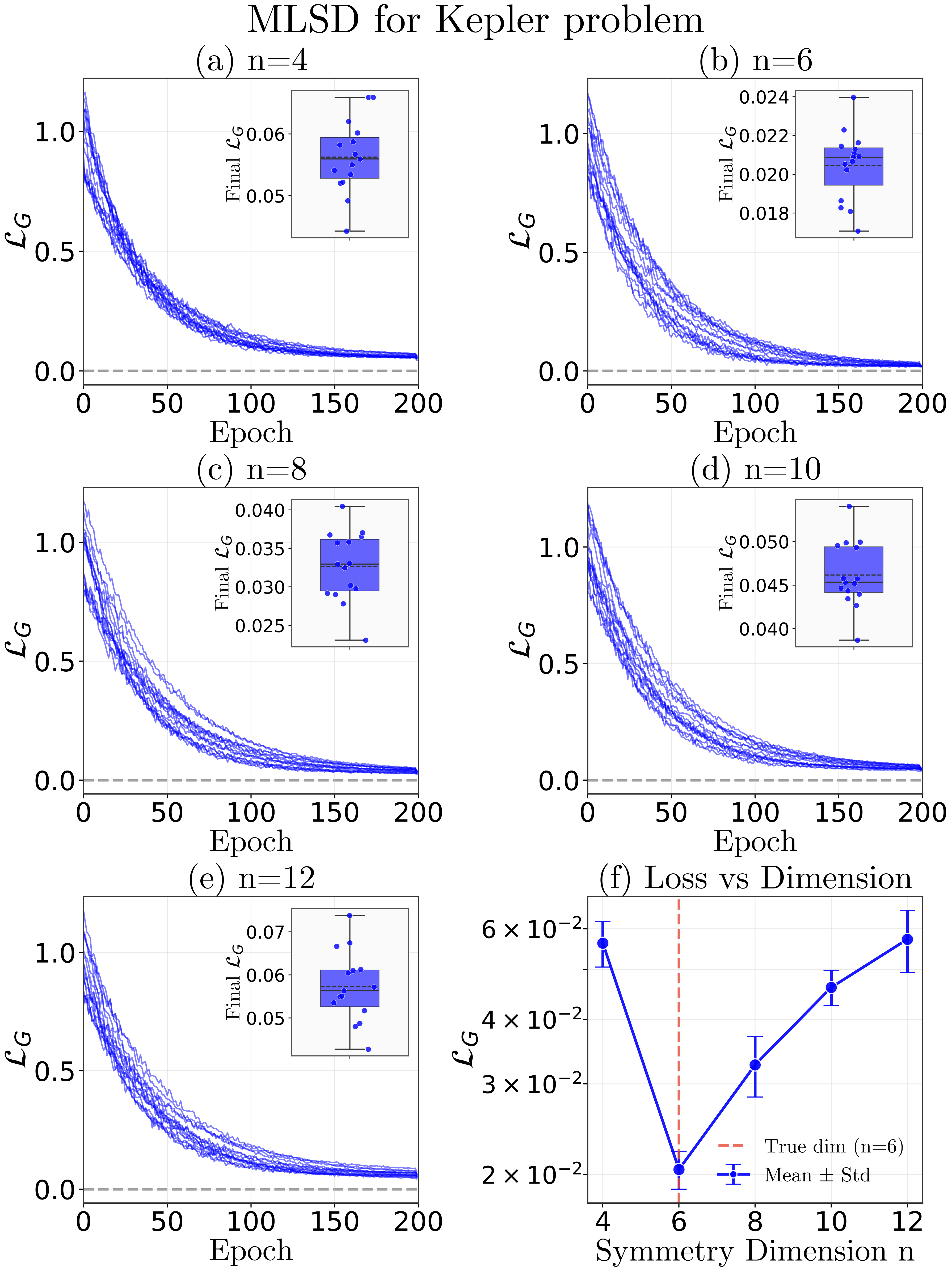}
\caption{MLSD for the 3D Kepler problem: training convergence and symmetry dimension identification.
    (a--e) Training loss $\mathcal{L}_G$ versus epoch for symmetry dimensions $n = 4, 6, 8, 10, 12$, 
    with 15 independent random seeds per dimension. 
    Insets show the distribution of final converged losses. 
    (f) Mean converged loss versus symmetry dimension $n$, with error bars indicating standard deviation across seeds. 
    The vertical dashed red line marks the true symmetry dimension $n = 6$, 
    which achieves the lowest loss, correctly identifying the $\mathfrak{so}(4)$ symmetry of the 3D Kepler problem.}
\label{fig: loss}
\end{center}
\end{figure}

Following the pipeline outlined in \secref{sec: method}, we tested multiple conjectured symmetry
dimensions $n=4, 6, 8, 10,$ and $12$.
For each choice of $n$, the neural networks were trained on the same simulation dataset using the
loss function in \eqref{eq: loss} as shown in \figref{fig: loss}. For each $n$, we repeat training with $15$ random seeds; error bars indicate the standard deviation across seeds. Among these, $n=6$ corresponds to the expected $\SO(4)$ symmetry of the Kepler problem, while
larger values such as $n=10$ represent over-parameterized hypotheses.
The remaining cases ($n = 4, 8, 12$) do not correspond to standard symmetry algebras and are included solely as complementary tests.

\begin{figure}[htbp]
\begin{center}
\includegraphics[width=240pt]{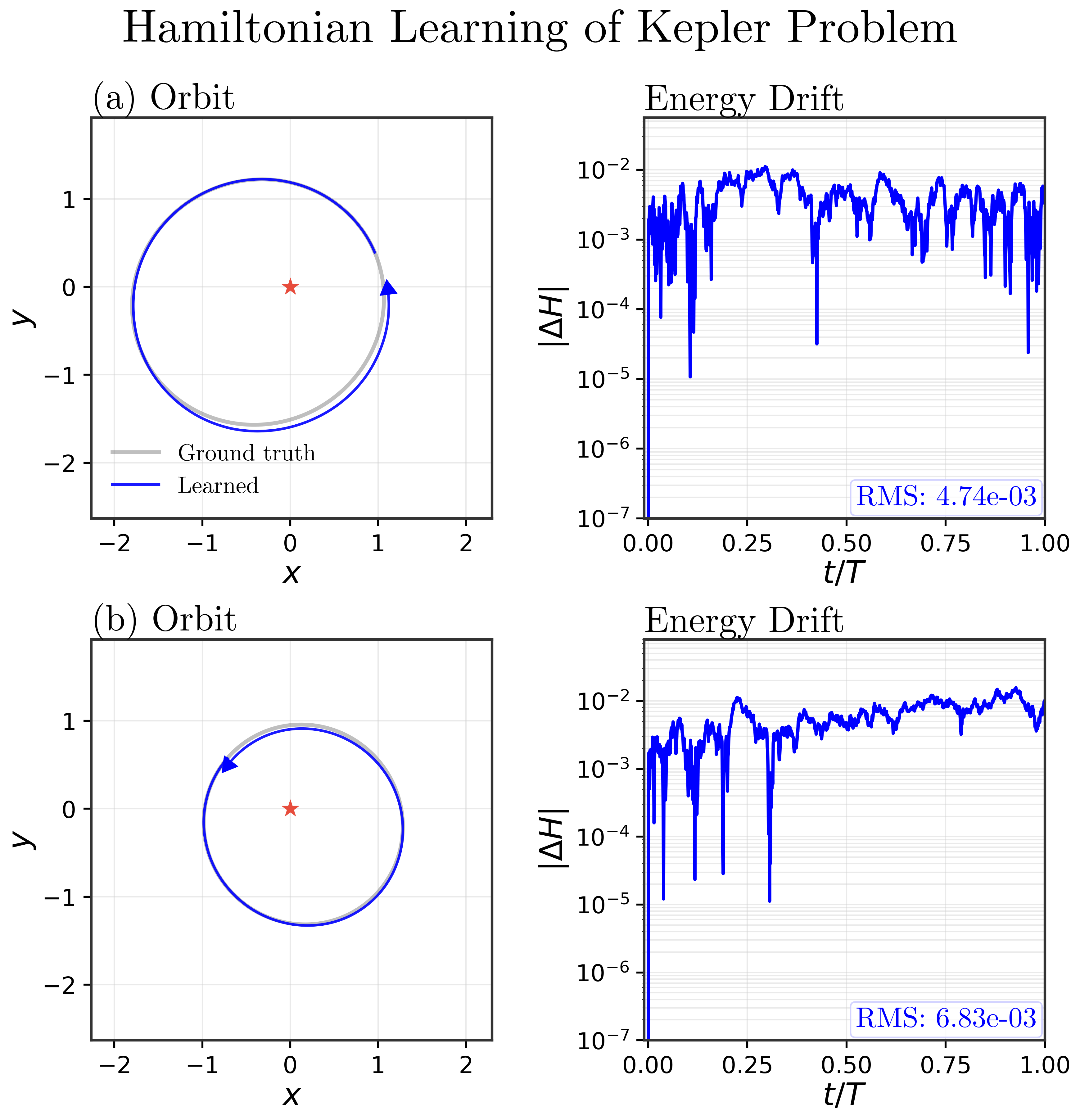}
\caption{Hamiltonian learning of the 3D Kepler problem: orbit comparison.
    Two representative examples (a, b) showing ground truth orbits (gray) versus learned trajectories (blue) integrated over one orbital period $T$. 
    The red star marks the central body at the origin. 
    Blue arrowheads indicate the velocity direction at the end of integration. 
    Right panels show the corresponding energy drift $|\Delta H|$ as a function of normalized time $t/T$, 
    demonstrating that errors accumulate due to imperfect Hamiltonian learning. All trajectories are integrated using the velocity-Verlet symplectic integrator.}
\label{fig: H}
\end{center}
\end{figure}

A straightforward way to assess the Hamiltonian learning is by comparing the resulting time-evolution trajectories, as illustrated in Fig.~\ref{fig: H}. Starting from the same initial condition $\vect{x}(t=0)$, the learned Hamiltonian $G_{0,\theta}$ reproduces the local time evolution generated by the true Hamiltonian $H$ with good agreement over short to intermediate times. Small deviations become apparent at longer times, which can be attributed to optimization
inaccuracies in the learned Hamiltonian $G_{0,\theta}$ together with the accumulation of local
numerical errors, even when using a symplectic integrator.
While these effects influence long-time trajectory accuracy, they remain tolerable for the present
tasks and do not obstruct the identification of the underlying symmetry structure.

Tracking the converged loss values in \eqnref{eq: loss} reveals a minimum at $n=6$ among the
tested dimensions $n=4,6,8,10,$ and $12$ (Fig.~\ref{fig: loss}).
While $\mathcal{L}_{\text{G}}$ does not vanish exactly due to optimization limitations, the marked
separation between $n=6$ and other cases indicates that the correct symmetry dimension is $n=6$.

To further identify the learned symmetry group, we analyze the machine-learned structure coefficients $f^{ijk}$ using two complementary post-processing procedures described in Appendix~\ref{app: killing}: (i) diagonalization of the Killing form and (ii) optimization of a linear transformation that maps the learned coefficients to a standard Lie-algebra basis. In practice, both approaches yield indistinguishable identification of the underlying algebra for the problems studied here.

\begin{figure}[htbp]
\begin{center}
\includegraphics[width=200pt]{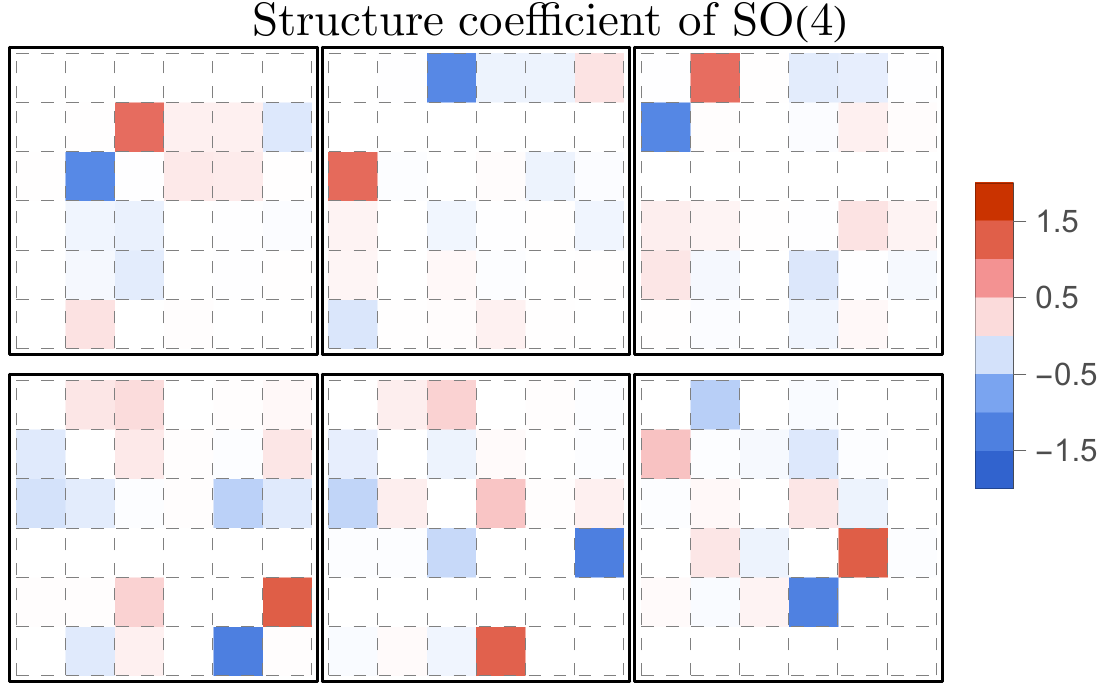}
\caption{The trained structure coefficient of Kepler problem.}
\label{fig: Kf}
\end{center}
\end{figure}

Specifically, we diagonalize the Killing form matrix $\mathbf{B} = \mathbf{U}\mathbf{D}\mathbf{U}^{T}$, whose entries are defined as $B^{il}=\sum_{k,j} f^{ijk}f^{lkj}$. A representative result yields $\mathbf{D}=\mathrm{diag}(-3.40,-3.44,-3.47)\oplus \mathrm{diag}(-1.22,-1.20,-1.17)$, which is consistent with the expected decoupling of two $\mathfrak{su}(2)$ subalgebras, since $\mathfrak{so}(4)\cong \mathfrak{su}(2)\oplus \mathfrak{su}(2)$. Further projecting the structure coefficients using $\mathbf{U}$ reproduces the standard $\mathfrak{so}(4)$ algebra, as shown in Fig.~\ref{fig: Kf}. For clarity, we present the Killing-form-based identification here, while both methods were applied for all reported results.

\subsection{Harmonic oscillator}

\begin{figure}[htbp]
\begin{center}
\includegraphics[width=240pt]{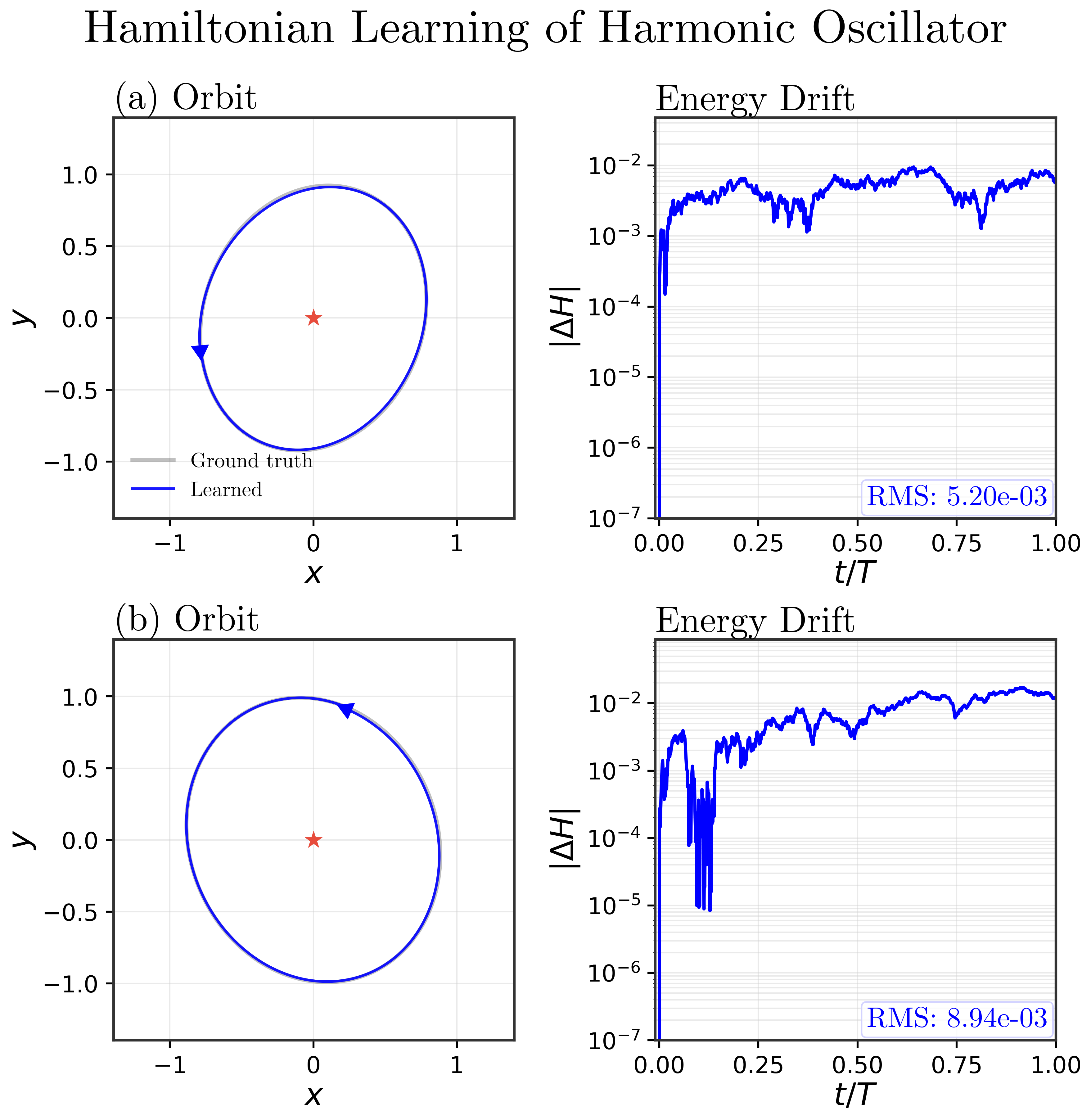}
\caption{Hamiltonian learning of the 3D harmonic oscillator: orbit comparison.
    Two representative examples (a, b) showing ground truth orbits (gray) versus learned trajectories (blue) integrated over one period $T$. 
    The red star marks the equilibrium position at the origin. 
    Blue arrowheads indicate the velocity direction at the end of integration. 
    Right panels show the corresponding energy drift $|\Delta H|$ as a function of normalized time $t/T$, 
    demonstrating error accumulation from imperfect Hamiltonian learning. All trajectories are integrated using the velocity-Verlet symplectic integrator.}
\label{fig: Hho}
\end{center}
\end{figure}

We apply the same procedure to the harmonic oscillator, testing symmetry dimensions
$n = 6, 7, 8, 9,$ and $10$. For each $n$, we repeat training with $15$ random seeds; error bars indicate the standard deviation across seeds.
The learned Hamiltonian dynamics (Fig.~\ref{fig: Hho}) show smaller long-time deviations than in the
Kepler case due to the simpler quadratic form of the harmonic oscillator Hamiltonian.

\begin{figure}[htbp]
\begin{center}
\includegraphics[width=220pt]{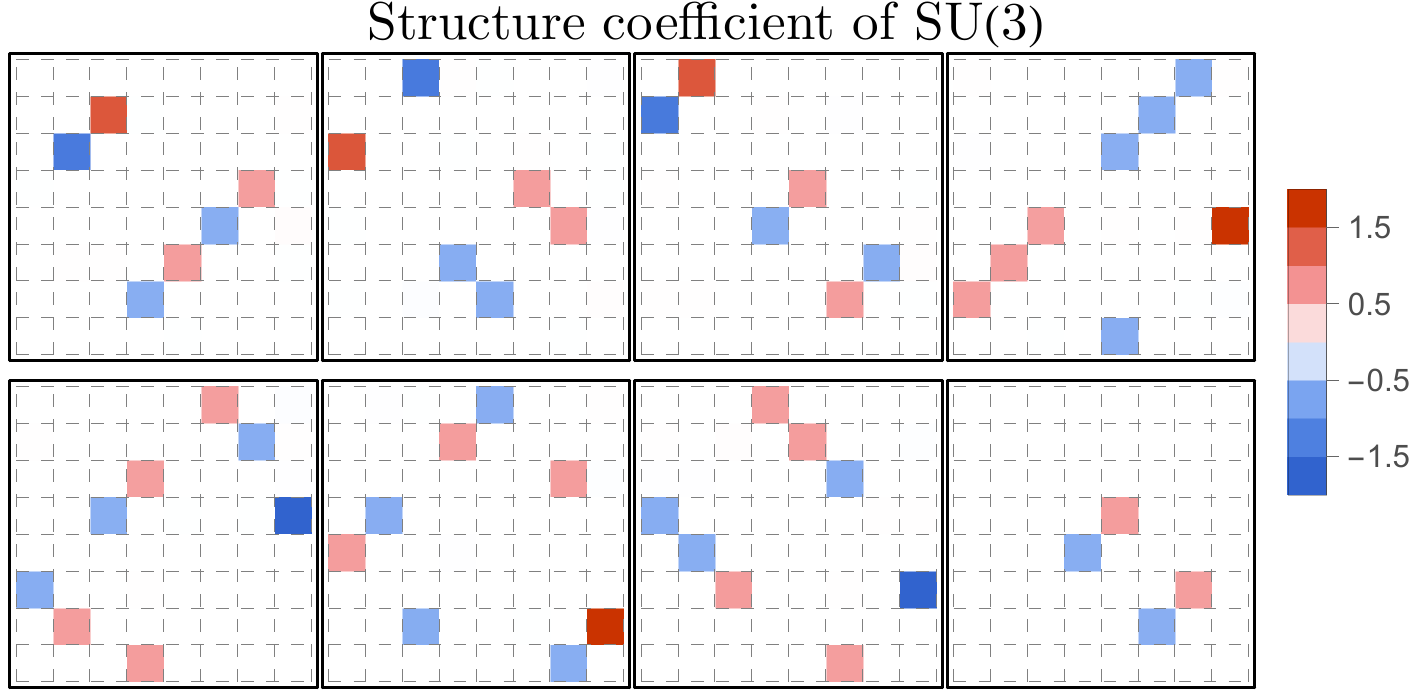}
\caption{The trained structure coefficient of harmonic oscillator, obtained through both quadratic and NN parametrizations, yield similar results that are identical to the $\su(3)$ Lie algebra.}
\label{fig: Hf}
\end{center}
\end{figure}

At $n=8$, the 8 eigenvalues of the Killing form of the structure coefficients exhibit strong degeneracy, for example: $\mathbf{D}=\mathrm{diag}$($-18.48$, $-18.46$, $-18.44$, $-18.43$, $-18.43$, $-18.41$, $-18.40$, $-18.38$). The structure coefficient reconstructs the $\su(3)$ Lie algebra as shown in the \figref{fig: Hf}. This strongly indicates a hidden  $\SU(3)$  symmetry in the harmonic oscillator system.

\begin{figure}[htbp]
\begin{center}
\includegraphics[width=240pt]{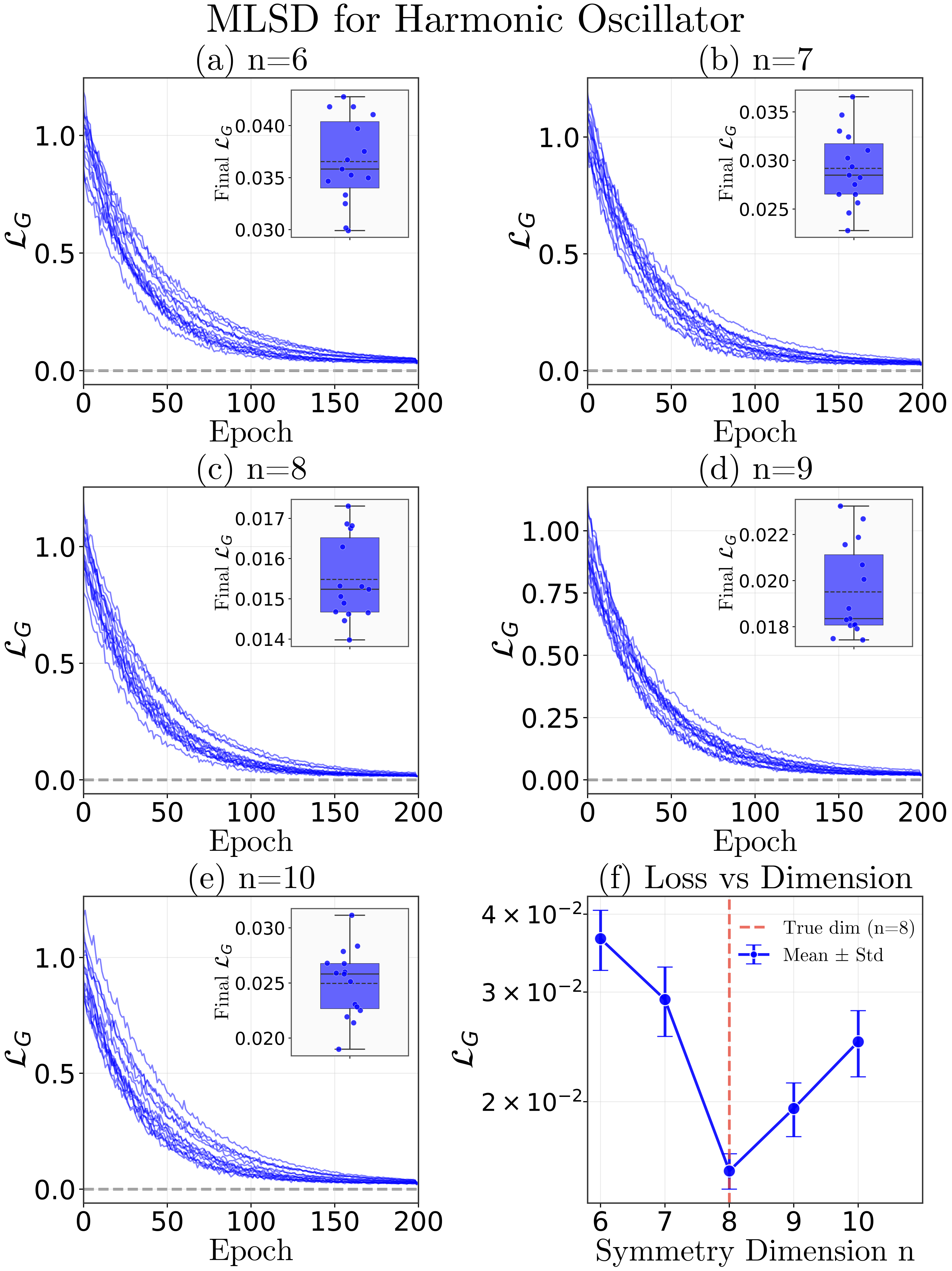}
\caption{MLSD for the 3D harmonic oscillator: training convergence and symmetry dimension identification.
    (a--e) Training loss $\mathcal{L}_G$ versus epoch for symmetry dimensions $n = 6, 7, 8, 9, 10$, 
    with 15 independent random seeds per dimension. 
    Insets show the distribution of final converged losses. 
    (f) Mean converged loss versus symmetry dimension $n$, with error bars indicating standard deviation across seeds. 
    The vertical dashed red line marks the true symmetry dimension $n = 8$, 
    which achieves the lowest loss, correctly identifying the $\mathfrak{su}(3)$ symmetry of the 3D harmonic oscillator.}
\label{fig: holoss}
\end{center}
\end{figure}

We also tried the method that parameterizes 9 variables in quadratic forms as $G_{i}=\vect{x}^TM_i\vect{x}$, where $i\in(0,1,2,...,8)$ and each $M_i$ is a real matrix of 6 by 6 that can be optimized. Here $G_0$ represents the Hamiltonian and others represent the 8 conserved quantities. 

\begin{figure}[htbp]
\begin{center}
\includegraphics[width=200pt]{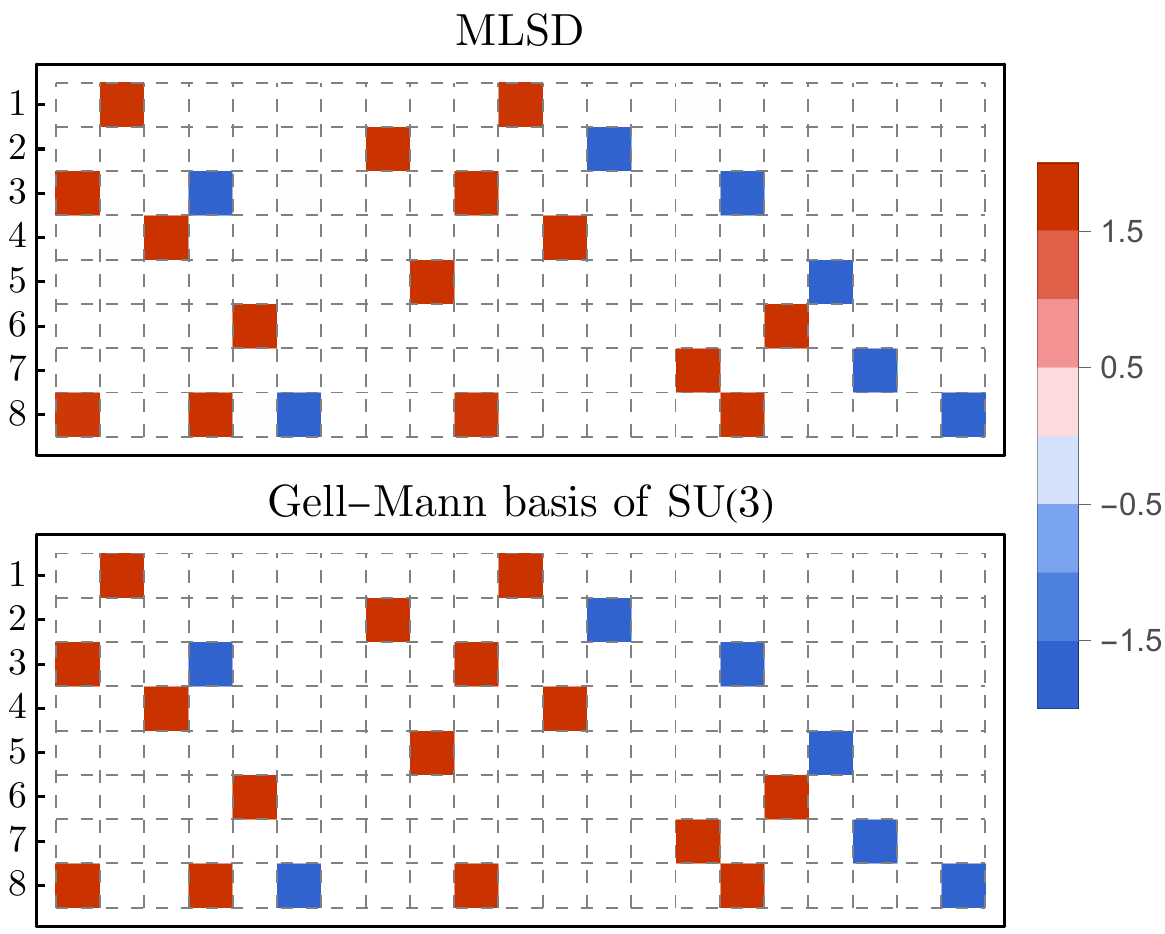}
\caption{Coefficient vectors of 8 quadratic quantities, each element represents the coefficient in the order of: $q_1q_1$, $q_1q_2$, $q_1q_3$, $q_2q_2$, $q_2q_3$, $q_3q_3$, $p_1q_1$, $p_1q_2$, $p_1q_3$, $p_1p_1$, $p_1p_2$, $p_1p_3$, $p_2q_1$, $p_2q_2$, $p_2q_3$, $p_2p_2$, $p_2p_3$, $p_3q_1$, $p_3q_2$, $p_3q_3$, $p_3p_3$.}
\label{fig: gell}
\end{center}
\end{figure}

After minimizing the loss function \eqref{eq: loss}, the 8 conserved quantities explicitly reconstruct the Gell-Mann basis \figref{fig: gell}.

\section{Summary and Discussion}\label{sec: summary}
\subsection{Symmetry discovery without prior knowledge}

In this study, we propose a data-driven, deep learning-based Machine Learning Symmetry Discovery algorithm designed to explore and analyze continuous symmetries in classical mechanical systems with integrable Hamiltonian dynamics. Previous works\cite{PhysRevLett.128.180201,PhysRevLett.126.180604,pmlr-v202-yang23n,2023arXiv231000105Y,otto2023unified,forestano2023deep,forestano2023identifying,2023PhRvE.108b2301Z} have demonstrated methods to verify conjectured continuous symmetries of a given Hamiltonian. Within this setting, our pipeline requires only the conjectured dimension of the symmetry group as input. The machine learning algorithm can then discover candidate symmetry transformations and subsequently identify the associated symmetry algebra, whether Abelian or non-Abelian, by analyzing the Killing form of the learned structure coefficients.

In a controlled benchmark setting, MLSD discovers candidate conserved quantities from trajectory data and recovers the associated Lie-algebraic structure via learned structure coefficients and Killing-form analysis. The contribution is a practical end-to-end pipeline, including reproducibility diagnostics across symmetry-dimension hypotheses, that enables systematic identification of non-Abelian symmetry algebras from data in integrable Hamiltonian systems.

\subsection{Choice of examples and limitations}

The present implementation of MLSD is constrained by the expressive capacity of the neural networks
used to parametrize $G_i(\vect{x})$.

The Kepler problem and the three-dimensional harmonic oscillator were chosen as demonstrations because they provide canonical benchmark systems and analytically well-understood symmetry structure, making them suitable test cases for validating the proposed framework. In systems with more complex or less regular dynamics, including mixed or chaotic phase space, the relevant invariants may involve higher-order nonlinearities, non-polynomial structure, or regime-dependent effective descriptions.

If the target symmetry transformations or the Hamiltonian $H(\vect{x})$ exceed the representational capacity of the chosen neural-network hypothesis class, the current implementation of MLSD may fail to recover the correct symmetry structure.

All results in this work are obtained using clean simulated trajectories expressed in canonical
coordinates.
The robustness of MLSD to measurement noise, partial observability, or imperfect coordinate
reconstruction is not addressed here.
Understanding how approximate symmetries degrade under noise and identifying noise thresholds
beyond which symmetry discovery fails are important and nontrivial directions for future work.

The MLSD algorithm also relies on an independence loss term $\mathcal{L}_{\text{I}}$ to encourage the
discovery of linearly independent symmetry generators.
Although weighted by a small parameter $\beta$, this term may introduce optimization sensitivity and
can affect convergence behavior.
In addition, symmetry-dimension selection in the present work is based on statistical comparison of
converged loss values across multiple random initializations rather than formal hypothesis testing.
While this provides a reproducible and practically effective criterion, developing more rigorous
model-selection or uncertainty-quantification frameworks is an important avenue for future
investigation.

Finally, the coverage of the training dataset plays a critical role in determining the symmetry structure recovered by MLSD. When trajectories sample diverse regions of phase space within a regular dynamical regime, the method is encouraged to recover symmetries of the global Hamiltonian. In contrast, for datasets restricted to narrow regions of phase space—including trajectories confined to KAM islands or chaotic subsets—the method may instead infer an effective, local symmetry algebra valid only within that region. Addressing these challenges through improved data-sampling strategies, more stable optimization
objectives, and more expressive architectures represents an important direction for future work.

\subsection{Extension to quantum and many-body system}

Extending the idea of symmetry discovery to quantum many-body systems is a highly valuable area of study. Understanding the hidden symmetries in these complex systems can provide deeper insights into the exotic properties of quantum states, and guide the classification of different states of matter based on their symmetries. However, directly computing symmetries in arbitrarily large quantum systems is infeasible due to the exponential growth of the Hilbert space. Therefore, developing a more efficient framework that combines renormalization group techniques\cite{2018NatPh..14..578K,hou2023machine,2022MLS&T...3c5009H} with symmetry discovery is an important avenue for future research. This approach could potentially reduce the computational complexity while still capturing the essential symmetries of the system, making it a promising direction for exploring the rich structures within quantum many-body physics.

\begin{acknowledgments}
We acknowledge helpful discussion with Rose Yu and Ziming Liu. WD and YZY are supported by the National Science Foundation (NSF) Grant DMR-2238360. ML is supported by UCSD Student Success Center 2024 Undergraduate Summer Research Award.

\end{acknowledgments}

\bibliographystyle{apsrev4-2}
\bibliography{ref}

\onecolumngrid
\newpage
\appendix
\setcounter{secnumdepth}{2}
\section{Symmetry Analysis}\label{app: sym}

\begin{itemize}
    \item Kepler problem

We begin by defining several fundamental quantities. Let the coordinates be \( \vect{q} = \{q_1, q_2, q_3\} \) and the momenta be \( \vect{p} = \{p_1, p_2, p_3\} \). The Hamiltonian is given by
\[
H = \frac{\vect{p}^2}{2} - \frac{1}{|\vect{q}|},
\]
the angular momentum vector is
\[
\vect{L} = \vect{q} \times \vect{p},
\]
and the Laplace–Runge–Lenz (LRL) vector is
\[
\vect{A} = \vect{p} \times \vect{L} - \frac{\vect{q}}{|\vect{q}|}.
\]

We utilize Poisson brackets to derive the commutation relations, defined as
\[
\{A, B\} = \frac{\partial A}{\partial q^i} \frac{\partial B}{\partial p_i} - \frac{\partial A}{\partial p^i} \frac{\partial B}{\partial q_i}.
\]
Calculate commutation relations between Hamiltonian:
\[
\{H, L_i\} = 0, \quad \{H, A_i\} = 0, \quad i \in \{1, 2, 3\}.
\]

This indicates that both \(\vect{L}\) and \(\vect{A}\) are conserved quantities. \\
Calculating the commutation relationships within and between \(\vect{L}\) and \(\vect{A}\):
\[
\{L_i, L_j\} = \epsilon^{ijk} L_k, \quad \{L_i, A_j\} = \epsilon^{ijk} A_k, \quad \{A_i, A_j\} = -2H \epsilon^{ijk} L_k, \quad ijk \in \{1, 2, 3\}.
\]

We define the scaled Laplace–Runge–Lenz vector
$\vect{N} \equiv \frac{\vect{A}}{\sqrt{-2H}}$,
and observe that \(\vect{N}\) and \(\vect{L}\) generate the \(\mathfrak{so}(4)\) algebra from the commutation relations above. Furthermore, we introduce the vectors
\[
\vect{\mathcal{L}} = \frac{1}{2} \left( \vect{L} + \vect{N} \right)
\]
and
\[
\vect{\mathcal{R}} = \frac{1}{2} \left( \vect{L} - \vect{N} \right).
\]

The commutation relations for \(\vect{\mathcal{L}}\) and \(\vect{\mathcal{R}}\) are given by:
\begin{align*}
\{\mathcal{L}_i, \mathcal{L}_j\} &= \epsilon^{ijk} \mathcal{L}_k, \\
\{\mathcal{R}_i, \mathcal{R}_j\} &= \epsilon^{ijk} \mathcal{R}_k, \\
\{\mathcal{L}_i, \mathcal{R}_j\} &= 0.
\end{align*}

These relations demonstrate that \(\vect{\mathcal{L}}\) and \(\vect{\mathcal{R}}\) each satisfy the commutation relations of the \(\mathfrak{su}(2)\). Consequently, we have two decoupled \(\mathfrak{su}(2)\), where
\[
\mathfrak{su}(2) \oplus \mathfrak{su}(2) \cong \mathfrak{so}(4).
\]

\item Harmonic oscillator

Using the same definitions of coordinate and momentum in the Kepler problem, we define the classical annihilation and creation functions as 
\[
\vect{a^{\dagger}} = \{a^*_1, a^*_2, a^*_3\} \quad \text{and} \quad \vect{a} = \{a_1, a_2, a_3\},
\]
where \( a_i = q_i + \mathrm{i} p_i, \; i \in \{1,2,3\} \).
\\
The Hamiltonian \( H \) is given by
\[
H = \frac{1}{2}(\vect{q}^2+\vect{p}^2) = \frac{1}{2}\vect{a^{\dagger}}\vect{a} = \frac{1}{2}\vect{a^{\dagger}} \mathds{1}\vect{a}.
\]
\\
Note that \(\{\vect{a^{\dagger}}A\vect{a}, \vect{a^{\dagger}}B\vect{a}\} = -2\mathrm{i}\vect{a^{\dagger}}[A,B]\vect{a}\). By constructing the generators as \( G_i = \frac{1}{2}\vect{a^{\dagger}}M_i\vect{a} \), we automatically have
\[
\{H, G_i\} = - \frac{\mathrm{i}}{2}\vect{a^{\dagger}}[\mathds{1},M_i]\vect{a} = 0.
\]
\\
We also want the structure constants to be anti-symmetric, leading to
\[
\{G_i, G_j\} = \frac{\mathrm{i}}{2}\vect{a^{\dagger}}[M_i, M_j]\vect{a} = f^{ijk}\frac{1}{2}\vect{a^{\dagger}}M_k\vect{a} = f^{ijk}G_k.
\]
\\
In this case, it is natural to choose \(\{M_i\}\) to be the Gell-Mann matrices, as the Gell-Mann matrices together with the identity matrix form a complete basis of 3 by 3 Hermitian matrices.
\\
The structure constants \( f^{ijk} \) turn out to be anti-symmetric, with the following non-zero values and their permutations:
\[
f^{123} = 2, \quad f^{147} = f^{246} = f^{257} = f^{345} = 1, \quad f^{156} = f^{367} = -1, \quad f^{456} = f^{678} = \sqrt{3}.
\]

\end{itemize}

\section{Basis transformation of Lie algebra}\label{app: killing}

\begin{itemize}
    \item Linear transformation of the structure coefficient

Considering two sets of Lie algebra basis $\vect{G}=(G_i,...,G_n)$ and $\vect{e}=(e_1,...,e_n)$ are related by a linear transformation by $M$:

\[
\vect{G} = M \vect{e}
\]
\[
\{e_a, e_b\} = \epsilon_{ab}^c e_c, \quad \forall a, b
\]
\[
\{G_i, G_j\} = f_{ij}^k G_k, \quad \forall i, j
\]
then the structure coefficient is related by
\[
f_{ij}^k M_k^c e_c = \{M_i^a e_a, M_j^b e_b\}
= M_i^a M_j^b \{e_a, e_b\}
= M_i^a M_j^b \epsilon_{ab}^c e_c
\]
\[
f_{ij}^k = M_i^a M_j^b \epsilon_{ab}^c (M^{-1})_c^k
\]
\[
(M^{-1})_a^i (M^{-1})_b^j f_{ij}^k M_k^c = \epsilon_{ab}^c
\]

Therefore, if there exists such a matrix $M$ that relates $f$ and $\epsilon$ through the above equality, then $\vect{G}$ and $\vect{e}$ can be identified as representations of the same Lie algebra.

\item Identify Lie group from diagonalizing the Killing form.

There are two methods to identify the Lie group after training with MLSD. Suppose  $f$  is the structure coefficient obtained from MLSD and  $\epsilon$  represents a reference coefficient. The straightforward approach is to use gradient descent to find an optimal transformation matrix  $M$  by minimizing the objective function: $\sum_{a,b,c}\|
(M^{-1})_a^i (M^{-1})_b^j f_{ij}^k M_k^c - \epsilon_{ab}^c\|$. If such a matrix $M$ exists, then the two groups are isomorphism to each other.

The second approach is to diagonalize the Killing form matrix defined as

$$\mathbf{B}_i^l=f^{k}_{ij}f^{j}_{lk}$$
$$\mathbf{D} = \mathbf{U}^{-1} \mathbf{B} \mathbf{U}$$

to obtain $\mathbf{D}$. The eigenvalues of the Killing form matrix  $\mathbf{B}$  will exhibit degeneracy corresponding to the sub-algebra structure of \( \vect{G} \). For instance, in the Kepler problem, the symmetry algebra is  $\mathfrak{so}(4) \cong \mathfrak{su}(2) \oplus \mathfrak{su}(2)$ . Consequently, diagonalizing the Killing form matrix of the trained structure coefficients  $f$  should yield two sets of degenerate eigenvalues: three identical eigenvalues corresponding to one $\mathfrak{su}(2)$ sub-algebra and another three identical eigenvalues corresponding to the other $\mathfrak{su}(2)$ sub-algebra. This eigenvalue pattern reflects the direct sum structure of the algebra. In the case of the harmonic oscillator problem, the symmetry algebra is  $\mathfrak{su}(3)$ , which cannot be decomposed into a direct sum of sub-algebras. Therefore, the eigenvalues of the Killing form matrix should exhibit a single set of eight identical values, reflecting the irreducibility of the  $\mathfrak{su}(3)$  algebra.
\end{itemize}

\end{document}